\documentclass[10pt,english,prb,amsmath,amssymb,prb,showkeys,superscriptaddress,twocolumn,showpacs,floatfix]{revtex4-1}
\usepackage{graphicx}
\usepackage{dcolumn}
\usepackage[colorlinks=true,linkcolor=blue ,citecolor=blue,urlcolor=blue]{hyperref}
\usepackage{multirow}
\usepackage[usenames,dvipsnames]{xcolor}
\usepackage{soul}
\usepackage{braket}
\usepackage{bm}
\usepackage{nicefrac}
\usepackage{siunitx}
\usepackage{color,transparent}
\usepackage[utf8]{inputenc}
\usepackage{upgreek}
\usepackage[export]{adjustbox}
\usepackage{pict2e}

\newcommand{\VEC}[1]{\mathbf{#1}}

\DeclareSIUnit\ML{ML}
\DeclareSIUnit\MLs{MLs}
\DeclareSIUnit\meVA{meV\angstrom^2}

\begin{document}

\title{An overview of the spin dynamics of antiferromagnetic Mn$_5$Si$_3$} 

\author{N. Biniskos}
\email{n.biniskos@fz-juelich.de}
\affiliation{Forschungszentrum J\"ulich GmbH, J\"ulich Centre for Neutron Science at MLZ, Lichtenbergstr. 1, D-85748 Garching, Germany}
\affiliation{Charles University, Faculty of Mathematics and Physics, Department of Condensed Matter Physics, Ke Karlovu 5, 121 16, Praha, Czech Republic}
\author{F. J. dos Santos}
\email{flaviano.dossantos@epfl.ch}
\affiliation{Laboratory for Materials Simulations, Paul Scherrer Institut, 5232 Villigen PSI, Switzerland}
\affiliation{Theory and Simulation of Materials (THEOS), and National Centre for Computational Design and Discovery of Novel Materials (MARVEL), \'Ecole Polytechnique F\'ed\'erale de Lausanne, 1015 Lausanne, Switzerland}
\author{M. dos Santos Dias}
\affiliation{Peter Gr\"unberg Institut and Institute for Advanced Simulation, Forschungszentrum J\"ulich $\&$ JARA, D-52425 J\"ulich, Germany}
\affiliation{Faculty of Physics, University of Duisburg-Essen and CENIDE, D-47053 Duisburg, Germany}
\affiliation{Scientific Computing Department, STFC Daresbury Laboratory, Warrington WA4 4AD, United Kingdom}
\author{S. Raymond}
\affiliation{Universit\'e Grenoble Alpes, CEA, IRIG, MEM, MDN, F-38000 Grenoble, France}
\author{K. Schmalzl}
\affiliation{Forschungszentrum J\"ulich GmbH, J\"ulich Centre for Neutron Science at ILL, 71 Avenue des Martyrs, F-38000 Grenoble, France}
\author{P. Steffens}
\affiliation{Institut Laue-Langevin, 71 avenue des Martyrs, 38000 Grenoble, France}
\author{J. Persson}
\affiliation{Forschungszentrum J\"ulich GmbH, J\"ulich Centre for Neutron Science (JCNS-2) and Peter Gr\"unberg Institut (PGI-4), JARA-FIT,  D-52425 J\"ulich, Germany}
\author{N. Marzari}
\affiliation{Theory and Simulation of Materials (THEOS), and National Centre for Computational Design and Discovery of Novel Materials (MARVEL), \'Ecole Polytechnique F\'ed\'erale de Lausanne, 1015 Lausanne, Switzerland}
\affiliation{Laboratory for Materials Simulations, Paul Scherrer Institut, 5232 Villigen PSI, Switzerland}
\author{S. Bl\"ugel}
\affiliation{Peter Gr\"unberg Institut and Institute for Advanced Simulation, Forschungszentrum J\"ulich $\&$ JARA, D-52425 J\"ulich, Germany}
\author{S. Lounis}
\affiliation{Peter Gr\"unberg Institut and Institute for Advanced Simulation, Forschungszentrum J\"ulich $\&$ JARA, D-52425 J\"ulich, Germany}
\affiliation{Faculty of Physics, University of Duisburg-Essen and CENIDE, D-47053 Duisburg, Germany}
\author{T. Br\"uckel}
\affiliation{Forschungszentrum J\"ulich GmbH, J\"ulich Centre for Neutron Science (JCNS-2) and Peter Gr\"unberg Institut (PGI-4), JARA-FIT, D-52425 J\"ulich, Germany}

\date{\today}

\begin{abstract}
The metallic compound Mn$_5$Si$_3$ hosts a series of antiferromagnetic phases which can be controlled by external stimuli such as temperature and magnetic field. 
In this work, we investigate the spin-excitation spectrum of bulk Mn$_5$Si$_3$ by combining inelastic neutron scattering measurements and density functional theory calculations. 
We study the evolution of the dynamical response under external parameters and demonstrate that the spin dynamics of each phase is robust against any combination of temperature and magnetic field. In particular, the high-energy spin dynamics is very characteristic of the different phases consisting of either spin waves or broad fluctuations patterns. 
\end{abstract}

\date{\today}

\maketitle

\section{Introduction}

Antiferromagnetic (AFM) compounds have recently emerged as a promising class of materials in the field of spintronics~\cite{Jungwirth2016,Baltz_2018}.
They exhibit a plethora of transport properties, such as the anomalous and the magnetic spin Hall effects discovered in Mn$_3$Sn~\cite{Nakatsuji2015,Kimata2019}.
As another example, the unusual tilted spin current discovered in RuO$_2$~\cite{Bose2022} is evidence for the newly-recognised class of altermagnetic materials~\cite{Smejkal_2022}.
These effects are already being exploited in multilayer structures that incorporate either heavy-metal or ferromagnetic layers, for instance to switch the magnetic order parameter using spin-orbit torques at terahertz (THz) frequencies~\cite{Olejnik2018}.

In fact, one of the attractive aspects of AFMs for spintronics applications is the much faster magnon dynamics in comparison with ferromagnets~\cite{rezende_introduction_2019}.
Magnons can also carry spin currents over long distances, as found in $\alpha$-Fe$_2$O$_3$~\cite{Lebrun2020} and in MnPS$_3$~\cite{Xing2019}.
In addition, magnons in AFM materials can display topological properties~\cite{diaz_topological_2019,dos_santos_modeling_2020} that parallel those found for electrons~\cite{Bonbien2021}, as exemplified by the Dirac magnons discovered in Cu$_3$TeO$_6$~\cite{Yao2018} and in CoTiO$_3$~\cite{Yuan2020}.
Therefore, combined theoretical and experimental investigations of magnons in AFM materials are essential to shed light on the fundamental physical effects and on how to adapt them to prospective spintronics devices.

In this article, we focus on the multifunctional AFM Mn$_{5}$Si$_{3}$.
This material exhibits interesting thermodynamic properties, such as the inverse magnetocaloric effect~\cite{biniskos_spin_2018} and thermomagnetic irreversibility~\cite{das_observation_2019}, and non-trivial transport properties through the anomalous Hall effect~\cite{surgers_large_2014}.
Hence, its macroscopic properties, crystal and magnetic structure have been extensively studied over the years~\cite{gottschilch_study_2012,brown_antiferromagnetism_1995,brown_low-temperature_1992,silva_magnetic_2002,biniskos_spin_2018,dos_Santos_2021,Biniskos_2022,das_observation_2019,surgers_large_2014,surgers_switching_2017}.
Moreover, a recent classification and description based on spin-symmetry principles proposes that Mn$_{5}$Si$_{3}$ belongs to a new distinct magnetic class called altermagnetism~\cite{Smejkal_2022}.
As it consists of cheap and abundant elements and can be grown as polycrystal~\cite{gottschilch_study_2012,das_observation_2019}, single crystal~\cite{Biniskos_2022,luccas_magnetic_2019}, thin film~\cite{surgers_large_2014,Kounta2023}, nanoparticle~\cite{das_mn5si3_2016}, and nanowire~\cite{sun_millimeters_2020}, it may attract further attention for the design and manufacturing of novel devices based on spintronic technologies.
Here, we microscopically probe the magnetization dynamics of Mn$_5$Si$_3$ in the meV regime (THz frequencies) utilizing inelastic neutron scattering (INS) measurements and density functional theory (DFT) calculations.
This combined approach is ideal for studying the spin-excitation spectrum under external parameters such as temperature ($T$) and magnetic field ($H$).  

\begin{figure}[t]
\setlength{\unitlength}{1cm} 
\newcommand{\boxsize}{0.3}
\begin{picture}(7.5,12.5)
    \put(0.3, 5.5){    
        \put(0., 0.3){ \includegraphics[width=3.5cm,trim={10cm 0 10cm 0},clip=true]{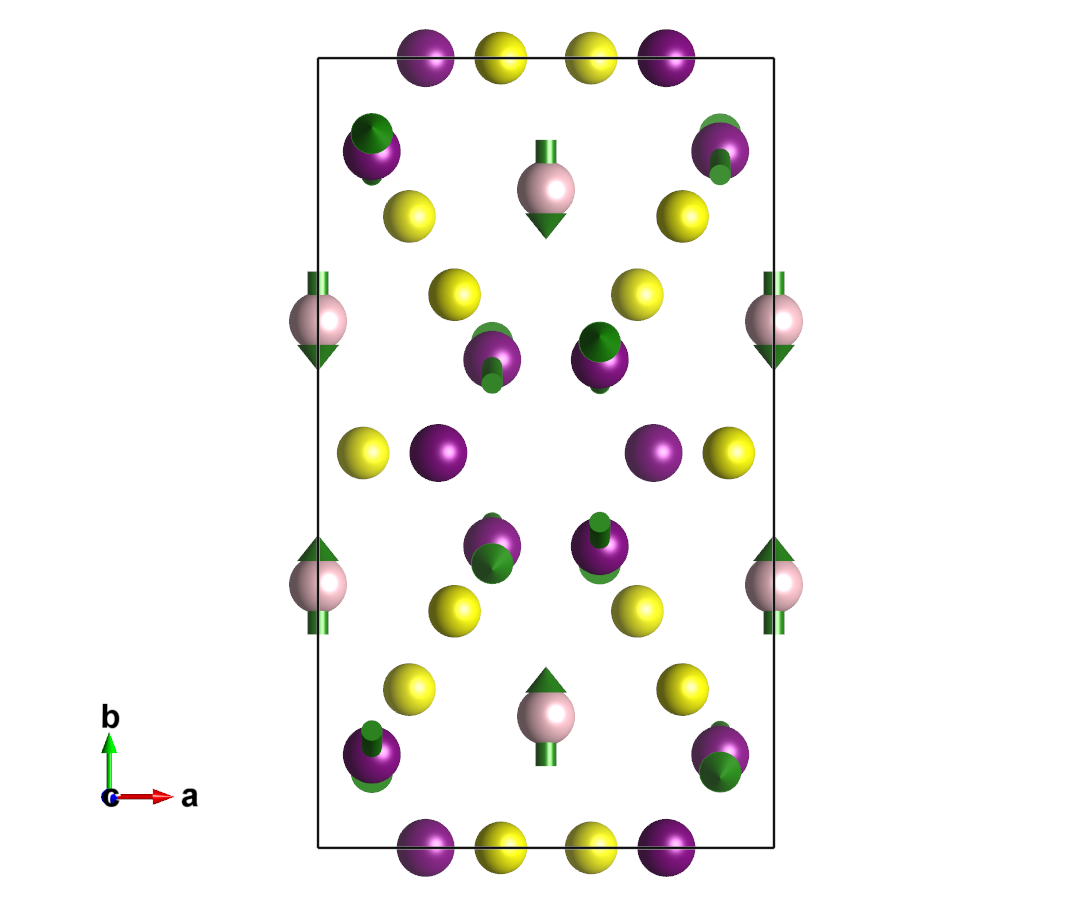}}
        \put(4.0, 0.3){ \includegraphics[width=3.5cm,trim={10cm 0 10cm 0},clip=true]{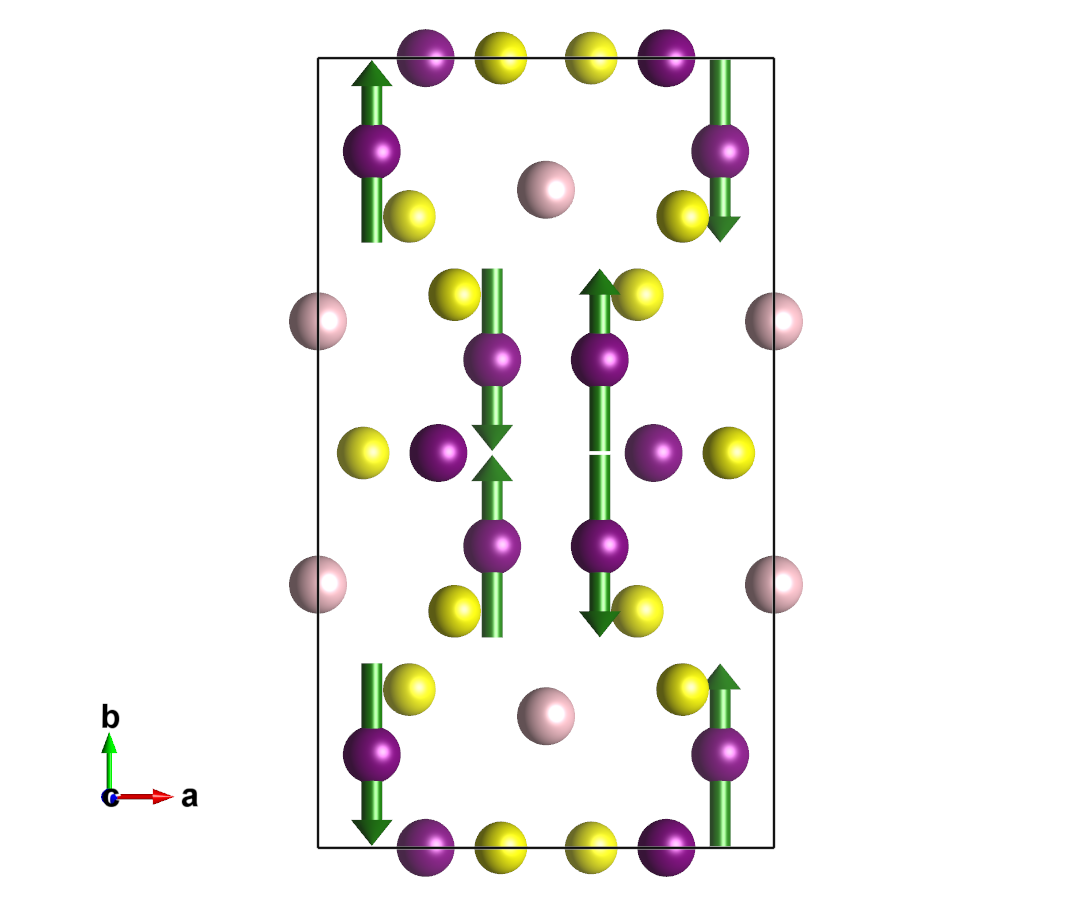}}
        
        \put(-0.3,  6.25){ \makebox(\boxsize,\boxsize){(a)} } 
        \put( 3.7,  6.25){ \makebox(\boxsize,\boxsize){(b)} } 

        \put( 1.25,  6.3 ){ \scalebox{1.1}{\makebox{AFM1}} }
        \put( 5.25,  6.3 ){ \scalebox{1.1}{\makebox{AFM2}} }
        
        \put(4.2, 3.4){
            \put( 0.80, 1.20){ \scalebox{1.0}{\makebox{Si}} }
            \put( 2.00, 0.15){ \scalebox{1.0}{\makebox{Mn2}} }
            \put( 1.20, 1.90){ \scalebox{1.0}{\makebox{Mn1}} } 
        }

        %Axes
        \put(-0.5,0.0) {
            \put( 0.05, 0 ){ \includegraphics[width=1.7cm,trim={30cm 29cm 10cm  10cm  },clip=true]{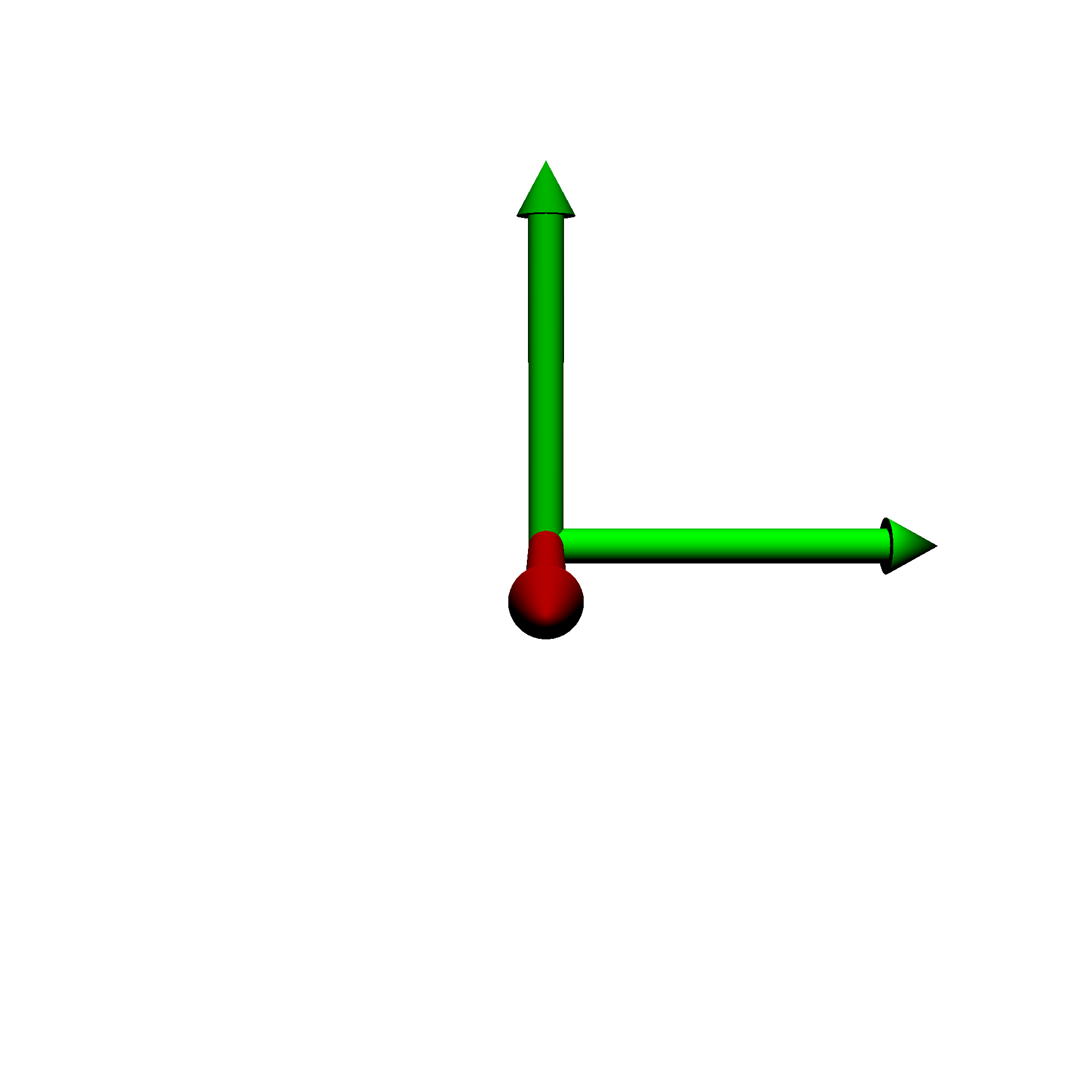} }
            \put( 1.8, 0.1){ \scalebox{0.9}{\makebox{a}} }
            \put(   0.1, 1.8){ \scalebox{0.9}{\makebox{b}} }
            \put(   0, 0.1){ \scalebox{0.9}{\makebox{c}} }
        }
    }
    \put(0.0, 0.0){    
        \put(0., 0.){ \includegraphics[width=7.5cm,trim={1cm 0 0 0},clip=true]{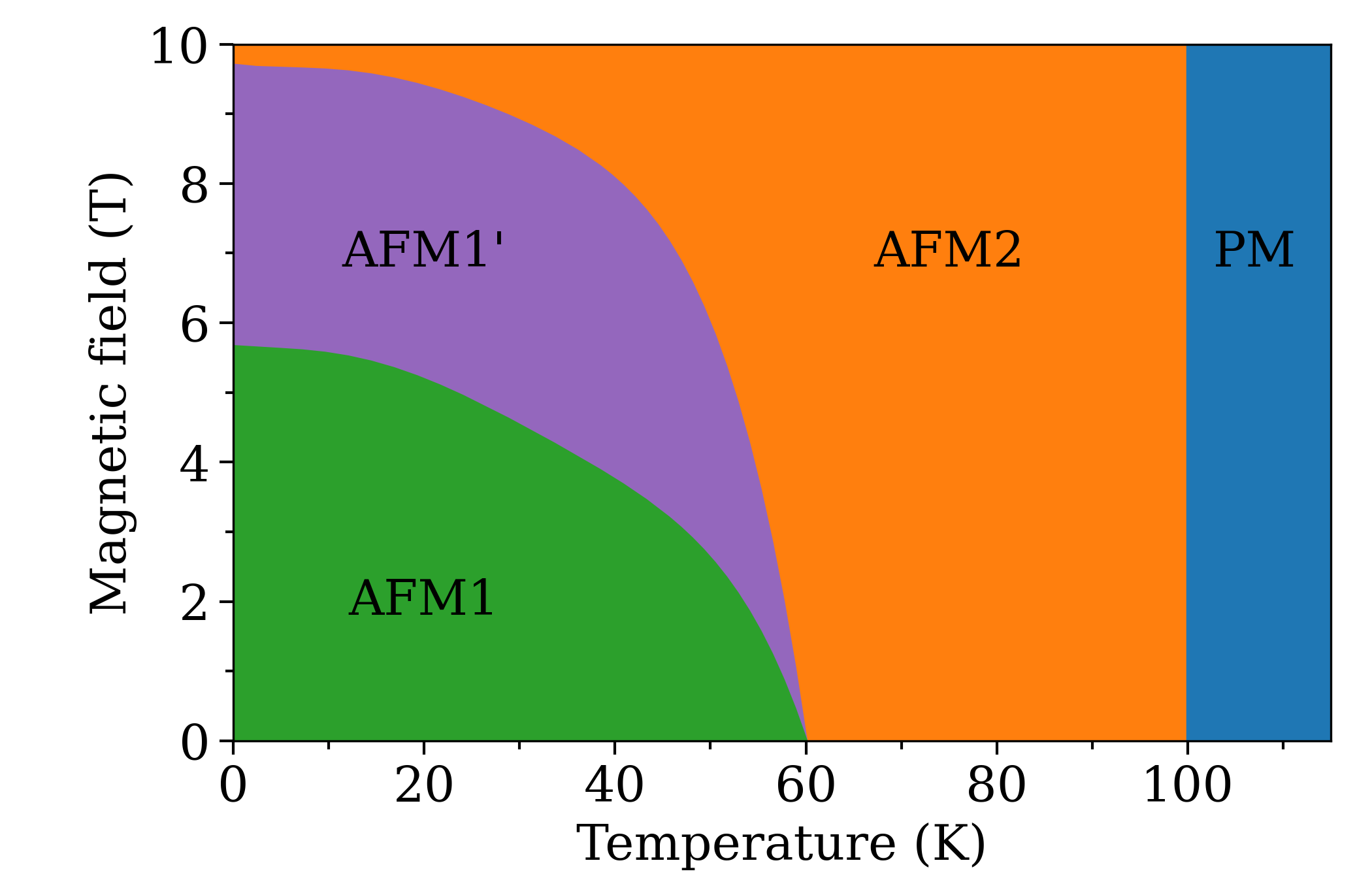}}
        \put(-0.4,  4.6){ \makebox(\boxsize,\boxsize){(c)} }
    }
\end{picture}
    \caption{
    Magnetic structures of bulk Mn$_{5}$Si$_{3}$ in the orthorhombic cell: (a) the noncollinear AFM1 phase~\cite{Biniskos_2022} and (b) the collinear AFM2 phase~\cite{brown_antiferromagnetism_1995,gottschilch_study_2012,dos_Santos_2021}. (c) Temperature and magnetic field phase diagram of Mn$_{5}$Si$_{3}$ for $\VEC H\parallel \hat{\VEC c}$ determined from macroscopic measurements (adapted from Ref.~\onlinecite{surgers_switching_2017}).
    \label{fig:all_AFM}
}
\end{figure}

Bulk Mn$_{5}$Si$_{3}$ is paramagnetic (PM) at $T > 100$\,K and crystallizes in the hexagonal space group $P6_{3}/mcm$ with two distinct crystallographic positions for the Mn atoms (sites Mn1 and Mn2)~\cite{gottschilch_study_2012}.
Below the PM critical temperature, the onset of a long-range magnetic order reduces the symmetry to orthorhombic~\cite{brown_antiferromagnetism_1995,brown_low-temperature_1992}.
Such a structural change is absent in epitaxially-grown thin films~\cite{Reichlova2020,Kounta2023}.
For temperatures between $66 < T < 100$\,K, the magnetic structure of Mn$_{5}$Si$_{3}$ is collinear (AFM2 phase) with only two-thirds of the Mn2 atoms having finite and ordered spins, which are aligned parallel and antiparallel to the $b$-axis.
The Mn1 atoms and the remaining one-third of the Mn2 atoms have vanishing or non-ordered moments~\cite{brown_antiferromagnetism_1995,gottschilch_study_2012,dos_Santos_2021} as shown in Fig.~\ref{fig:all_AFM}(b). 
For even lower temperatures ($< 66$\,K), the system realizes a noncollinear spin structure (AFM1 phase), where the Mn1 spins are now ordered but one-third of Mn2 spins remains as in the AFM2 phase.
The noncollinearity is a result of frustration and anisotropy, where the Mn1 spins align mostly along the easy axis $b$, while the two-thirds of the Mn2 spins have significant components along the $c$-axis as illustrated in Fig.~\ref{fig:all_AFM}(a)~\cite{Biniskos_2022}.

Both AFM phases are so-called mixed magnetic phases~\cite{Lacroix_2010} embodying the coexistence of magnetically ordered and non-ordered Mn sites~\cite{gottschilch_study_2012,brown_antiferromagnetism_1995,brown_low-temperature_1992}.
The magnetic phase diagram as a function of temperature and magnetic field applied along the $c$-axis~\cite{surgers_switching_2017} is shown in Fig.~\ref{fig:all_AFM}(c).
Below $T = 66$\,K, the increasing magnetic field results in transitions from the AFM1 phase to another intermediate AFM1' phase before reaching the AFM2 phase.
For the temperature range $66 < T < 100$\,K, the AFM2 phase remains stable up to the investigated field of 10\,T.
We also note that apart from the transitions occurring in the $H-T$ phase diagram~\cite{surgers_switching_2017,das_observation_2019} and presented in Fig.~\ref{fig:all_AFM}(c), there are indications of another AFM phase~\cite{gottschilch_study_2012,surgers_switching_2017,AitHaddouch_2022} appearing at small magnetic fields at low temperatures ($H < 1$\,T, $T < 30$\,K) and surviving in a narrow magnetic field range~\cite{Biniskos_2022} but that is outside the scope of this work.

\section{Previous results}

For our studies, we used the orthorhombic coordinate system and the scattering vector $\VEC Q$ is expressed in Cartesian coordinates $\VEC Q = (Q_{h}, Q_{k}, Q_{l})$ given in reciprocal lattice units (r.l.u.). The wave vector $\VEC q$ is related to the momentum transfer through $\hbar\VEC Q = \hbar\VEC G + \hbar\VEC q$, where $\VEC G$ is an AFM zone center and $\VEC G = (h, k, l)$.
Several aspects of the magnetism of Mn$_{5}$Si$_{3}$ were addressed in our previous works by combining INS measurements and DFT calculation~\cite{biniskos_spin_2018,dos_Santos_2021,Biniskos_2022}.
The ground state spin configuration (AFM1 phase) and the different magnetic structures induced at low temperatures under magnetic field were theoretically determined~\cite{Biniskos_2022}. 
The spin-wave spectra were studied in detail in the AFM1~\cite{Biniskos_2022} ($T=10$\,K) and AFM2~\cite{dos_Santos_2021} ($T=80$\,K) phases both theoretically and experimentally. 
These zero-field spin dynamics results will be summarized in Section IV together with new results obtained for the spin-wave spectra for the phases reached under magnetic field (AFM1' and field-induced AFM2). 

In addition, an insight on how the high-energy spin dynamics (above $\approx$ 2\,meV) evolves between the AFM1 and AFM2 phase was given in Ref.~\onlinecite{biniskos_spin_2018}.
It was shown that the spin dynamics in the AFM2 phase resembles that of the PM state: the response is broad in $\VEC Q$ with a single peak centred at the AFM zone center $\VEC G$ in contrast to the spin-wave spectrum of the AFM1 phase having two peaks on both sides of $\VEC G$. 
In Ref.~\onlinecite{biniskos_spin_2018}, the peculiar spectrum of the AFM2 phase was analyzed as arising from the coexistence of spin waves and spin fluctuations given the fact that the response obtained by polarized INS experiment is distinct between the AFM2 and PM states. 
This interpretation was made using the simplest possible hypothesis but it cannot be ruled out that a more complex dynamical response occurs. 
Therefore, in the present study, we will name the signal which is characteristic of the AFM2 phase above $\sim$ 2\,meV as ``broad fluctuations''. 
Its physical origin is ascribed to non-magnetic Mn sites occurring within the magnetically ordered phase. 
The inverse magnetocaloric effect (cooling by adiabatic magnetization) was related to this change of spin dynamics from well-defined spin waves to broad fluctuations under field~\cite{biniskos_spin_2018}. 

The present paper is organized as follows: after recalling briefly the methods that are similar to the previous papers~\cite{biniskos_spin_2018,dos_Santos_2021,Biniskos_2022} (Section III), an overview of the spin configuration and spin-wave spectra is given for all phases of the $H-T$ phase diagram in Section IV.A. 
Then, a focus on the temperature and magnetic field dependence of the spin dynamics is made at the low (Section IV.B) and high-energy (Section IV.C) regimes. 

\section{Methods}

A Mn$_{5}$Si$_{3}$ single crystal (the same used in previous studies~\cite{biniskos_spin_2018,dos_Santos_2021,Biniskos_2022}) with mass of about 7\,g and grown by the Czochralski method was oriented in the [100]/[010] scattering plane of the orthorhombic symmetry.
Unpolarized INS measurements were carried out on the cold triple-axis spectrometers (TAS) IN12~\cite{schmalzl_upgrade_2016} and ThALES at the Institut Laue-Langevin (ILL) in Grenoble, France. 
The instrument configurations are standard and are given in the previous papers~\cite{biniskos_spin_2018,dos_Santos_2021,Biniskos_2022}. 
Inelastic scans were performed with constant $k_f = 1.5$\,{\AA}$^{-1}$, where $\VEC{k}_f$ is the wave vector of the scattered neutron beam.
Spin dynamics investigations under temperature and magnetic field were carried out using a 10\,T vertical field magnet.

The magnetic parameters were computed with the DFT code juKKR~\cite{Bauer2014,jukkr}, which are then used to solve a spin Hamiltonian in the linear spin-wave approximation~\cite{dosSantos2018,dos_santos_first-principles_2017}.
The Monte-Carlo Metropolis algorithm as implemented in the Vampire package~\cite{evans_atomistic_2014,vampire} was employed.
A macro cell with 1\,$\mu$m$^3$ was considered with equilibration time of $10^5$ Monte-Carlo steps, and statistical averaging over $10^5$ steps.

\section{Results}

\subsection{Spin configurations and spin-wave dispersions}

\begin{figure}[t]
    \includegraphics[width=8.5 cm]{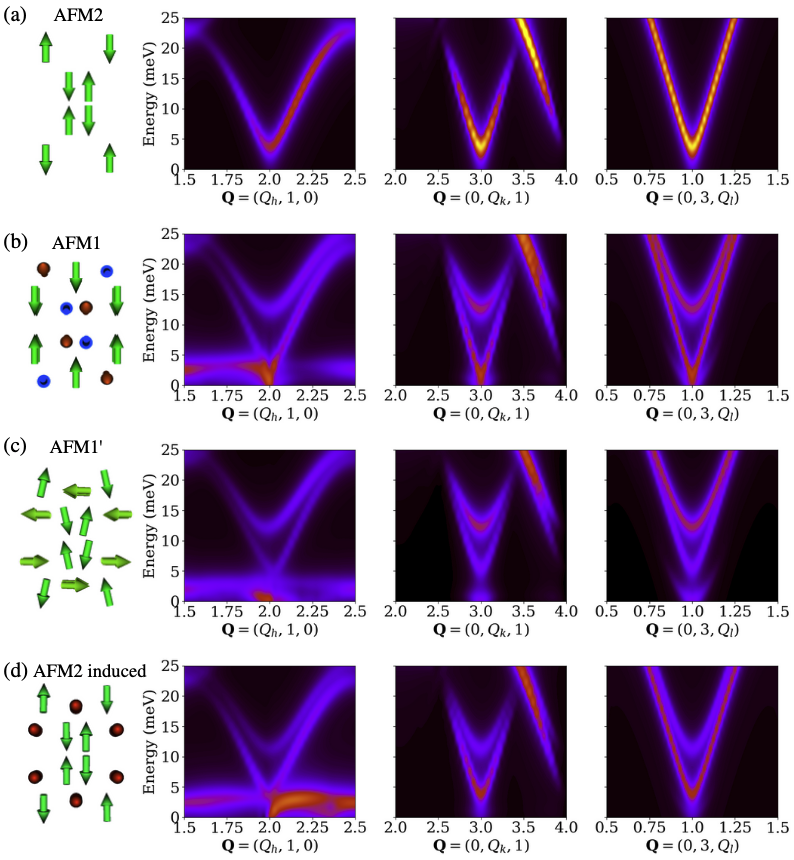}
    \caption{
    Spin configuration (based on Ref.~\onlinecite{Biniskos_2022}) and simulated spin-wave dispersion relations of Mn$_{5}$Si$_{3}$ at the (a) AFM2, (b) AFM1, (c) AFM1', and (d) field-induced AFM2 phases. In each phase the spin orientation of the Mn atoms having magnetic moment is shown as arrows. The spin structure is projected in the $ab$-plane of the orthorhombic cell. The color map corresponds to the calculated magnon spectrum along the high symmetry directions ($h$00), (0$k$0), and (00$l$).}
    \label{fig:dft}
\end{figure}

Figure~\ref{fig:dft} shows for each AFM phase of Mn$_{5}$Si$_{3}$ the computed spin structure and the corresponding magnon dispersion relations originating from the purely magnetic Bragg peaks (2,1,0) and (0,3,1). 
The AFM2 [Fig.~\ref{fig:dft}~(a)] and AFM1 [Fig.~\ref{fig:dft}~(b)] phases were previously addressed in Refs.~\onlinecite{dos_Santos_2021} and \onlinecite{Biniskos_2022}, respectively, and in the present paper, we extend the results to the AFM1' and field-induced AFM2 phases.
The DFT calculations~\cite{dos_Santos_2021} for the zero-field AFM2 phase [Fig.~\ref{fig:dft}~(a)] reproduce the experimental magnetic structure determined by neutron diffraction~\cite{brown_antiferromagnetism_1995}.  
As previously mentioned, this phase has the peculiarity that only two-thirds of the Mn2 sites have ordered magnetic moments.
At temperatures below 60\,K, the Mn1 sites acquire a finite magnetic moment which leads to a destabilization of the collinear AFM2 ordering.
In DFT, a finite moment at the Mn1 sites is achieved in a spin polarized calculation~\cite{Biniskos_2022}.
At zero field, we have the AFM1 phase [Fig.~\ref{fig:dft}~(b)], which is noncollinear due to the frustrated antiferromagnetic coupling between the Mn1 and Mn2 moments.
An applied magnetic field can induce phase transitions into the AFM1' [Fig.~\ref{fig:dft}~(c)] and the so called field-induced AFM2 [Fig.~\ref{fig:dft}~(d)] phases, where the moment at the Mn1 site is still finite.

We point out that there is a degree of uncertainty in the obtained magnetic interactions, which affects the phase boundaries between the different AFM phases shown in Fig.~\ref{fig:all_AFM}(c).
Therefore, the quantitative disagreement~\cite{Biniskos_2022} between the experimental and the theoretical phase boundaries in the $H-T$ phase diagram can probably be ascribed to the uncertainties in the parameterization of the spin model.
However, the PM critical temperature of the model Hamiltonian is 110\,K, as determined from Monte Carlo simulations, which is in good agreement with the measured transition temperature of 100\,K~\cite{dos_Santos_2021,das_observation_2019}.
This suggests that the obtained exchange interactions from our calculations are in the correct energy scale of the system.

In contrast to Ref.~\onlinecite{dos_Santos_2021} that dealt uniquely with the AFM2 phase, in the modeled spin-wave spectrum the Hamiltonian parameters are not rescaled to account for thermal fluctuations for the zero-field collinear AFM2 phase observed experimentally at $66 \leq T \leq 100$\,K [see Fig.~\ref{fig:all_AFM}(c)].
In the calculated spin waves, we observe for $E \geq 2$\,meV similar characteristics in the magnon bands of the AFM1, AFM1' and field-induced AFM2 phases despite their differences in the magnetic structure.
All these phases share the same Hamiltonian (exchange couplings and anisotropy).
The difference is that the applied field causes a change in the ground-state spin configuration but only small changes in the high energy magnon spectra.
This seems to indicate that the spin configuration does not affect strongly the spin-wave properties as the interactions between spins do.

Another feature that becomes apparent is that the spectrum of the zero-field AFM2 phase consists only of an acoustic mode in the energy range of $2 \leq E \leq 25$\,meV, as can be seen in Fig.\ref{fig:dft}(a).
In contrast, all the other phases feature optic bands originating from the AFM zone centers at about $E = 12$\,meV.
This difference can be explained by the vanishing of the magnetic moments in the Mn1 sites in the zero-field AFM2 phase.
Thus, the optic mode is mainly due to the AFM coupling between spins in the Mn1 and Mn2 sites ($|J| \sim 2$\,meV~\cite{Biniskos_2022}).

If we now focus on the low-energy spin dynamics ($E < 2$\,meV), we recall that there is a single anisotropy gap of about 0.8\,meV in the AFM1 phase~\cite{Biniskos_2022} and a double gap of about 0.2\,meV and 0.4\,meV in the AFM2 phase~\cite{dos_Santos_2021} ascribed to the bi-axial anisotropy [see Fig.~\ref{fig:gap}(b)].
We also note that although Mn atoms with non-ordered moments are present in all phases, the effects associated with fluctuating moments are difficult to model theoretically.
However, their fingerprint can be detected through neutron scattering experiments (see Sections IV.B and C).

\subsection{Low-energy spin dynamics ($E < 2$\,meV)}

\begin{figure}[ht]
    \includegraphics[width=8 cm]{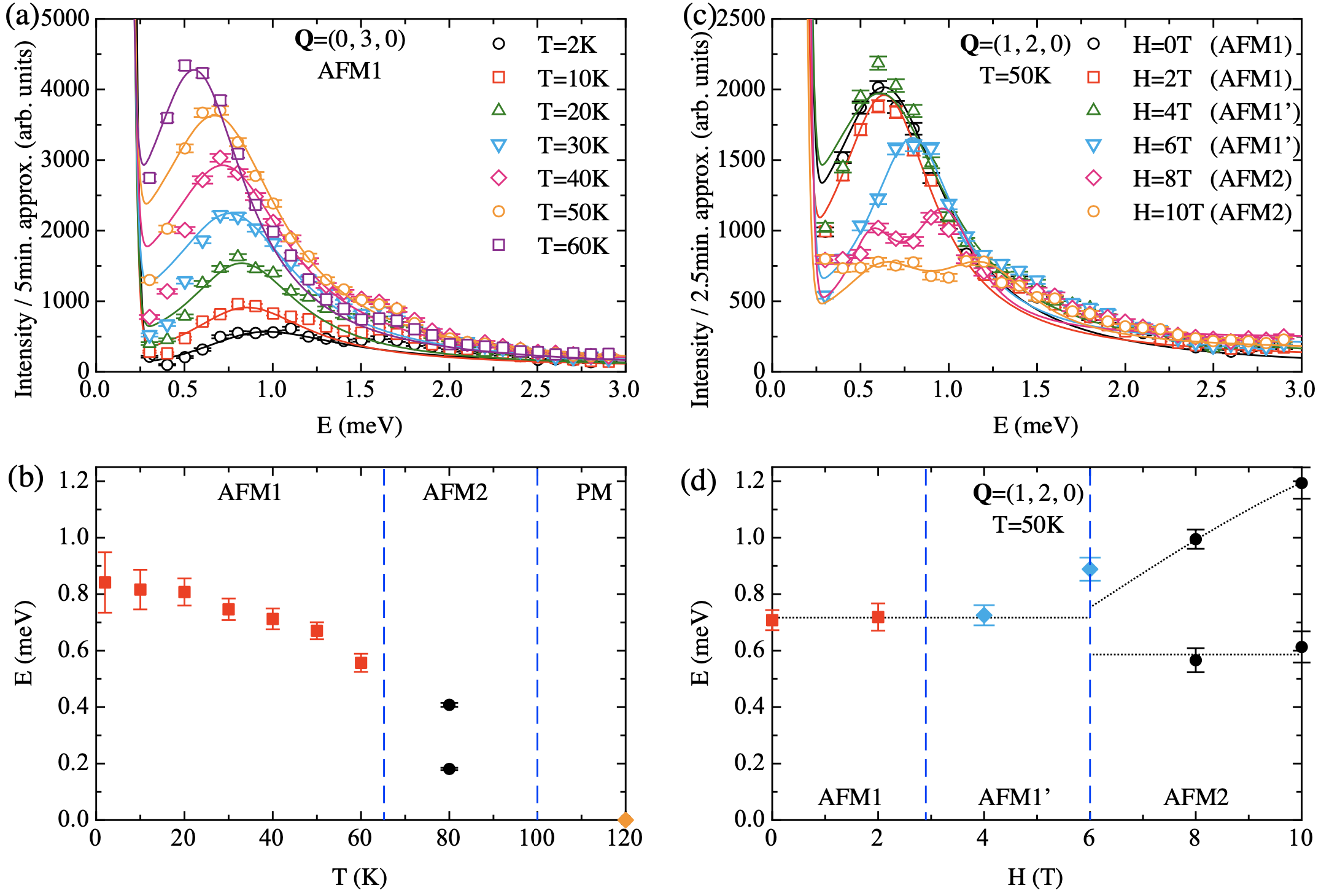}
    \caption{
    Low-energy spin dynamics of Mn$_{5}$Si$_{3}$ under temperature and magnetic field ($\VEC H\parallel \hat{\VEC c}$). 
    (a) Temperature dependence of constant-$\VEC Q$ scans collected at $\VEC Q = (0, 3, 0)$ in the AFM1 phase at $H = 0$\,T. 
    (b) Temperature dependence of the spin gap at $H = 0$\,T. The data points at $T = 80$\,K (AFM2 phase) are taken from Ref.~\onlinecite{dos_Santos_2021}.
    (c) Field dependence of constant-$\VEC Q$ scans collected at $\VEC Q = (1, 2, 0)$ at $T = 50$\,K. 
    (d) Field dependence of the spin gap at $T = 50$\,K.
    Dashed vertical blue lines in (b) and (d) separate the different AFM phases. In (d) the black dashed lines are guides for the eyes.
    \label{fig:gap}
}
\end{figure}

Starting from the AFM1 phase, the temperature and magnetic field dependence of the low-energy spin dynamics ($E < 2$\,meV) was not addressed in our previous studies. 
This is done here through INS measurements around the magnetic zone centers $\VEC G = (0, 3, 0)$ and $\VEC G = (1, 2, 0)$ [see Figs.~\ref{fig:gap}]. 
Fig.~\ref{fig:gap}(a) shows the temperature dependence of the spin excitations obtained at $H = 0$\,T.
A first peak is always centered at $E = 0$\,meV and corresponds to the elastic line, while the second one at finite $E$ points to the existence of gapped spin waves.
In the constant-$\VEC Q$ scans a sharp rise in the signal at $0.3 < E < 1.2$\,meV is followed by a large tail extending to high energies [see Figs.~\ref{fig:gap}(a) and (c)].
This tail could arise from resolution effects in relation to the steep magnon dispersion in the AFM1 phase~\cite{Biniskos_2022}.
Another scenario is that spin fluctuations originating from non-ordered moments would result to a broad quasielastic contribution extending to high energies and being overlaid over the signal of well-defined spin waves.

Within the resolution of our measurements, we observe for the AFM1 phase a single gap which is opening with decreasing temperature [see red data points in Fig.~\ref{fig:gap}(b)]. 
This spin gap hints to a local easy axis existing within this noncollinear phase and our DFT calculations indicate that the easy axis of the AFM1 phase is $b$.
Note that in the AFM2 phase, the primary and secondary easy axes are $b$ and $c$, respectively~\cite{dos_Santos_2021}, resulting to a double spin gap [see black data points at $T = 80$\,K in Fig.~\ref{fig:gap}(b)]. 
The higher energy gap of the AFM2 phase is solely determined by the $b$-axis anisotropy parameter, while the lower energy gap involves the difference of anisotropy parameters between the $b$ and $c$-axes~\cite{dos_Santos_2021}. 
Consistently, one can see in Fig.~\ref{fig:gap}(b) that the temperature dependence of the AFM1 gap extrapolates to the value of the higher energy gap of the AFM2 phase.
Naturally, no spin gap is observed in the PM state since spin fluctuations due to short range correlations result in a continuum of states [see orange data point at $T = 120$\,K in Fig.~\ref{fig:gap}(b)].

By applying a field parallel to the $c$-axis in the AFM1 phase, we induce phase transitions [see Fig.~\ref{fig:all_AFM}(c)] that also manifest in the low-energy spin dynamics. 
Spectra obtained at $\VEC G = (1, 2, 0)$ at $T = 50$\,K as a function of field are shown in Fig.~\ref{fig:gap}(c). 
First, weak magnetic fields ($H \leq 2$\,T) do not affect the gap in the AFM1 phase, however, for higher fields ($H \geq 6$\,T) the gap increases and splits recovering a double-gapped structure characteristic of the zero-field AFM2 phase [see $T = 80$\,K in  Fig.~\ref{fig:gap}(b)], but here for the field-induced AFM2 phase ($T = 50$\,K and $H = 8$ and 10\,T). 
The obtained field dependence of the spin gaps at $T = 50$\,K is shown in  Fig.~\ref{fig:gap}(d). 
While the AFM1 structure is complex with components of ordered moments parallel and perpendicular to the applied field, the gap does not changes between 0 and 2\,T. 
The value of the gap in the middle of the AFM1’ phase (at $H = 4$\,T) is similar to the one in the AFM1 phase and increases at 6\,T in the boundary with the field-induced AFM2 phase.
In contrast, a clear variation can be addressed for the field-induced AFM2 phase. 
Starting from the AFM1 phase at 50\,K and applying a magnetic field, the behaviour is the same as for the 80\,K AFM2 phase under field for the same field range 6 to 10\,T: the (zero field) higher energy mode does not depend on the magnetic field while the (zero field) lower one increases monotonically with field [see Fig. 4 of Ref.~\onlinecite{dos_Santos_2021}; note also that the modes cross at 4\,T].
In this former paper (Ref.~\onlinecite{dos_Santos_2021}), the field response of the spin waves in the AFM2 phase at 80\,K was theoretically explained by considering the precession directions of the different magnon modes.

\begin{figure}[t]
    \includegraphics[width=8 cm]{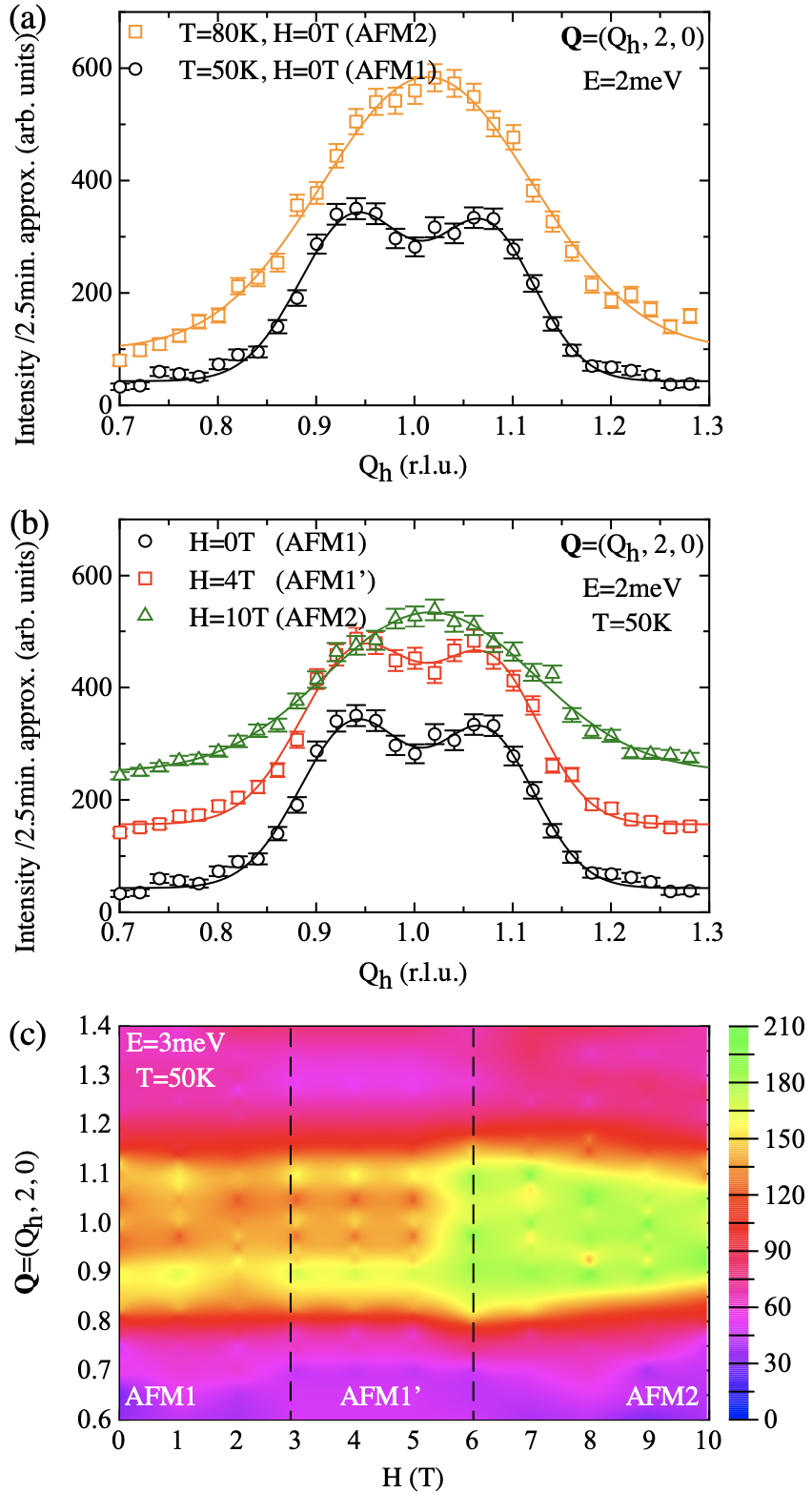}
    \caption{
    Inelastic spectra at constant energy transfers at $\VEC Q = (Q_{h}, 2, 0)$ in the different AFM phases of Mn$_{5}$Si$_{3}$.
    (a) INS data collected at $E = 2$\,meV in the AFM2 ($T = 80$\,K) and AFM1 ($T = 50$\,K) phases at $H = 0$\,T.
    (b) Field dependence ($\VEC H\parallel \hat{\VEC c}$) of spectra at $T = 50$\,K collected at $E = 2$\,meV.
    The spectra are shifted for clarity in intensity conserving the same scale.
    (c) Color-coded intensity plot of spectra collected at $E = 3$\,meV at $T = 50$\,K as a function of magnetic field.
    Lines in (a) and (b) indicate fits with Gaussian functions.
    The vertical lines in (c) separate the different AFM phases.
    \label{fig:Map1}
}
\end{figure}

\subsection{High-energy spin dynamics ($E \geq 2$\,meV)}

Fig.~\ref{fig:Map1}(a) shows the evolution of constant energy spectra collected along $\VEC Q = (Q_{h}, 2, 0)$ at 2\,meV for $T = $\,50 and 80\,K at zero field. 
It is similar to data previously obtained at slightly higher energies of 3 and 5\,meV~\cite{biniskos_spin_2018} and illustrates the change in the spin dynamics from two separate spin-wave peaks ($T = 50$\,K) to broad fluctuations ($T = 80$\,K) as described in Section II. 
A similar change in the spectra can also be obtained at 50\,K as a function of field as shown in Fig.~\ref{fig:Map1}(b) for 0 and 10\,T. 
Since the temperature is now fixed in this measurement, this indicates that the origin of the fluctuations is not a thermal broadening but that they are characteristic of the AFM2 phase. 

The overall dynamical response is also depicted in Fig.~\ref{fig:Map1}(c) in a color-coded map of the INS intensity measured for another constant energy transfer of $E = 3$\,meV at $T = 50$\,K as a function of $\VEC Q = (Q_{h}, 2, 0)$ and magnetic field. 
As seen in Figs.~\ref{fig:Map1}(b) and ~\ref{fig:Map1}(c), at $T = 50$\,K as the external magnetic field increases within the AFM1 and AFM1' phases the two peaks remain resolved and unchanged (a consistent behavior with the calculated spectra shown in Figs.~\ref{fig:dft}), until the field-induced AFM2 phase is reached. Then, the signal is substituted by a broad fluctuation pattern. 
In the color-coded map [Fig.~\ref{fig:Map1}(c)], this manifests through the change in the spectra from two intensity ridges at $Q_{h} \simeq 0.9$\,r.l.u. and $Q_{h} \simeq 1.1$\,r.l.u. to one intensity ridge centered at $Q_{h} = 1$\,r.l.u.. 

\begin{figure}[t]
    \includegraphics[width=8 cm]{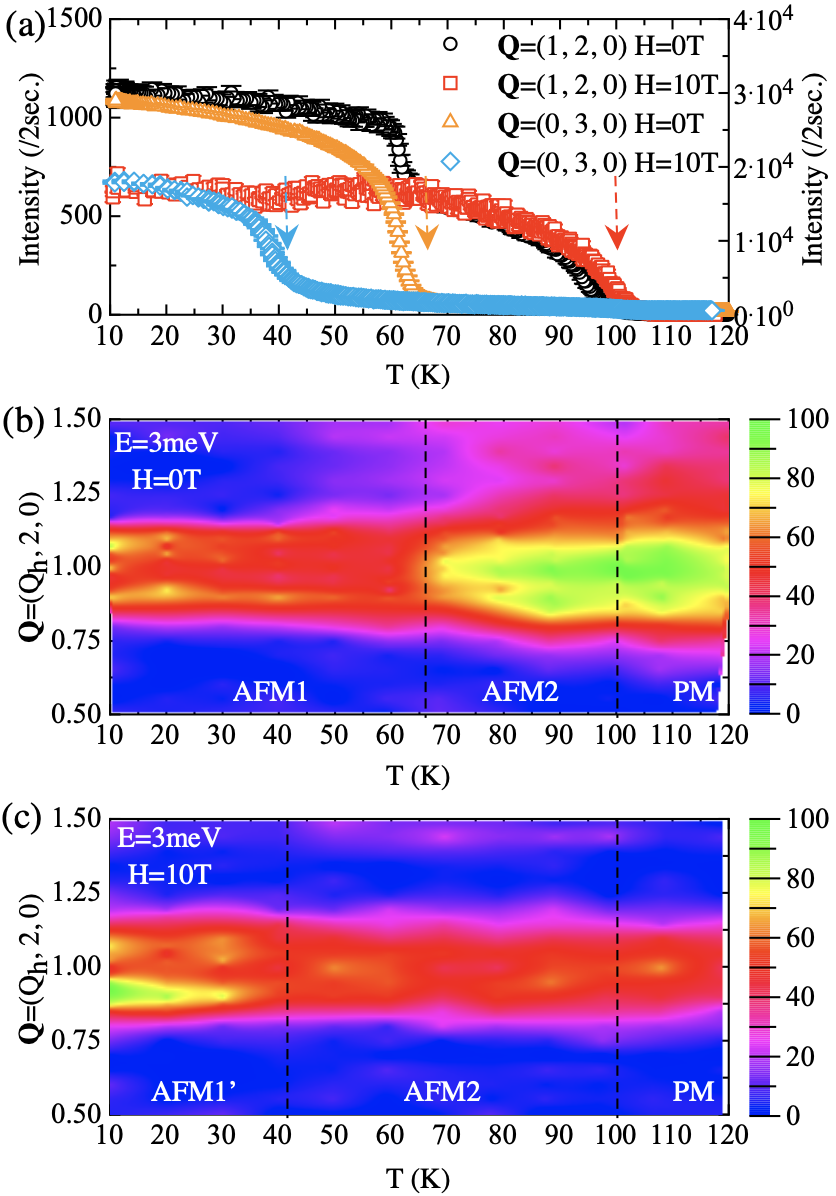}
    \caption{
    Temperature and magnetic field ($\VEC H\parallel \hat{\VEC c}$) dependence of (a) magnetic (1, 2, 0) and (0, 3, 0) Bragg peak intensities, and color-coded intensity plot of the INS data collected at $E = 3$\,meV as a function of $\VEC Q = (Q_{h}, 2, 0)$ at (b) $H = 0$\,T, and (c) $H = 10$\,T.
    The intensities for (1, 2, 0) and (0, 3, 0) are given on the left and right vertical axis, respectively. 
    For (1, 2, 0), a 10\,mm plexiglass attenuator was in place.
    In the inelastic spectra, a constant background is subtracted and the measured intensity is corrected by the detailed balance factor.
    The arrows in (a) mark the onset of the AFM transitions.
    The vertical lines in (b) and (c) separate the different AFM phases.
    \label{fig:Map2}
}
\end{figure}

In order to further support our statements, data on the temperature dependence of the spin-excitation spectrum of Mn$_{5}$Si$_{3}$ at 0 and 10\,T are shown in Figs.~\ref{fig:Map2}, where the magnetic Bragg peaks intensities are used to track the phase boundaries in the $H-T$ diagram. 
The temperature dependence of the peak intensities at $\VEC G = (1, 2, 0)$ and $\VEC G = (0, 3, 0)$ at $H = 0$\,T and $H = 10$\,T is shown in Fig.~\ref{fig:Map2}(a). 
At $H = 0$\,T, with reducing temperature, the abrupt increase of intensity for the (1, 2, 0) Bragg peak first at $T = 100$\,K and then at $T = 66$\,K marks the onset of the AFM ordering, in good agreement with the phase diagram shown in Fig.~\ref{fig:all_AFM}(c). 
The (0, 3, 0) reflection is absent in the AFM2 phase~\cite{brown_low-temperature_1992} and is used for confirming the AFM1 (or AFM1') ordering. 
At $H = 10$\,T, the transition temperature of the AFM2 phase, as seen from the (1, 2, 0) peak intensity, is weakly affected.  
The (0, 3, 0) peak intensity indicates that a phase transition occurs at about 40\,K (AFM2 $\rightarrow$ AFM1'), which is in agreement with a small intensity change observed for (1, 2, 0) in the same temperature. 
For $T < 40$\,K and $H = 10$\,T, Mn$_{5}$Si$_{3}$ is still in the AFM1' phase and magnetic fields higher than 10\,T may be needed to reach the field-induced AFM2 phase at low temperatures (we recall that the construction of the phase diagram shown in Fig.~\ref{fig:all_AFM}(c) is based on macroscopic measurements).
Our neutron data corroborate previous results about the field-induced 
transitions in Mn$_{5}$Si$_{3}$~\cite{das_observation_2019,gottschilch_study_2012,silva_magnetic_2002,surgers_switching_2017,AitHaddouch_2022,Alkanani_1995}. 

The corresponding changes for the inelastic response measured along $\VEC Q = (Q_{h}, 2, 0)$ at 3\,meV are shown in Figs.~\ref{fig:Map2}(b) and ~\ref{fig:Map2}(c).
At zero field, the switch in the spin dynamics manifests with increasing temperature as previously described from the change of two spin-wave intensity ridges to a single ridge centered at $Q_h = 1$\,r.l.u. that resembles that of the PM state ($T > 100$\,K).
At 0\,T, this occurs at 66\,K (AFM1 $\rightarrow$ AFM2 transition), while at 10\,T, the switch between the AFM1' and field-induced AFM2 phases occurs around 40\,K.
Therefore, the broad fluctuations are now observed in a wider temperature range from 45 to 100\,K stressing again that they do not correspond to a thermal broadening of the spin waves of the AFM1 (or AFM1') phase but that this excitation spectrum is an intrinsic feature of the AFM2 phase. 
Altogether, the change of the spin dynamics between AFM1 (or AFM1') and AFM2 is observed along three different paths in the $H-T$ phase diagram: temperature dependance at 0\,T [Fig.~\ref{fig:Map2}(b)], field dependence at 50\,K [Fig.~\ref{fig:Map1}(c)] and temperature dependence at 10\,T [Fig.~\ref{fig:Map2}(c)].

\section{Discussion}

In Mn$_{5}$Si$_{3}$, the peculiar excitation spectrum of the AFM2 phase is naturally associated with the presence of ordered and non-ordered Mn sites.
This is based on the observation that it looks at first sight similar to the PM state, but it also involves spin waves as demonstrated by polarized INS experiments in Ref.~\onlinecite{biniskos_spin_2018}. 
This response could be a complex entanglement between spin waves and spin fluctuations beyond the simple sum proposed in Ref.~\onlinecite{biniskos_spin_2018} and, therefore, in the present paper we named it broad fluctuations. 
Whatever its precise nature is, one concludes that the different features of the spin dynamics characteristic of the AFM1 or AFM2 phases are linked to changes of the magnetic behaviour of specific Mn sites as a function of temperature and magnetic field. 

At present, it is unclear if the observed fluctuations arise from the Mn1 or the one-third non-ordered Mn2 sites.
Our theoretical results so far do not reproduce the broad features of the spin-excitation spectra of the zero-field and field-induced AFM2 phases [see in Fig.~\ref{fig:dft} top and bottom panels] since fluctuating moments (spins changing their sizes and directions) are at the present stage hard to characterize from DFT calculations. 
Our model indicates the presence of magnetic moments on the Mn1 site in the field-induced AFM2 phase [see Fig.~\ref{fig:dft} bottom left].
However, based on a three-spin model introduced in Ref.~\onlinecite{Biniskos_2022}, it was demonstrated that a non-rigid Mn1 moment in a frustrated environment can fluctuate and its size change depends on the competition of the exchange interactions and the magnetic field strength. 
Concerning the one-third of Mn2 sites in the field-induced AFM2 phase, a neutron diffraction study proposes that there is no induced moment under external magnetic field~\cite{silva_magnetic_2002}, however, neutron scattering is not directly a site selective probe and further hypotheses must be made when discussing such aspects. 
Last but not least, one needs to consider that the dynamical response of the non-ordered moments may manifest in other ($\VEC Q$, $E$) regions that were not investigated in our INS study either in the AFM1 (AFM1') or AFM2 phases.

The dynamics of the fluctuations within the mixed magnetic phases of Mn$_{5}$Si$_{3}$ could be potentially studied by various experimental techniques probing different energy scales like a.c.\ susceptibility, neutron spin echo, nuclear magnetic resonance (NMR) and muon-spin rotation ($\mu\textnormal{SR}$). 
Particularly local probes, such as NMR~\cite{Panissod_1984} and $\mu\textnormal{SR}$, can be employed to investigate the magnetization dynamics at specific crystallographic sites and obtain information that is not accessible by neutron experiments.
Apart from Mn$_{5}$Si$_{3}$, mixed magnetic phases are found in several metallic and insulating magnetic compounds~\cite{Ballou_1991,Ballou_1996,Lacroix_1996,Ehlers_2001,Oyamada_2008,Rossat_1983,Vogt_1995}. 
The possible consequences of the fluctuating or non-ordered moments in the physical properties of interest for spintronics is not yet addressed to our knowledge.

Finally, we mention that besides the impacts of temperature and magnetic field, an increasing amount of evidence points out that pressure could significantly affect the magnetic structure and consequently the magnon spectrum of Mn$_{5}$Si$_{3}$. 
Magnetization measurements under hydrostatic pressure indicate a destabilization of the AFM1 phase with increasing pressure leading eventually to ferromagnetic ordering~\cite{Vinokurova1995}.
Chemical pressure induced by carbon implantation~\cite{Suergers2009}, by doping the Mn-site~\cite{songlin_magnetic_2002,Das2021,Das2022} or the Si-site~\cite{Haug1979,Zhao_2006} is demonstrated to have drastic changes on the electric and magnetic properties of Mn$_{5}$Si$_{3}$.
The inherent stress Mn$_{5}$Si$_{3}$ single crystals acquire when grown by different techniques is proposed to lead into deviations on the $H-T$ phase diagram shown in Fig.~\ref{fig:all_AFM}(c)~\cite{luccas_magnetic_2019}.
Lastly, epitaxial strain may potentially stabilize the AFM2 ordering to a wider temperature range~\cite{Reichlova2020}.
Note that for these very thin films, it is proposed from the measured physical properties that the AFM2 ordering is of different kind than in the bulk material and occurs without a doubling of the unit cell~\cite{Reichlova2020}.
Therefore, a dedicated study on the effects of uniaxial or hydrostatic pressure on the spin structure of bulk Mn$_{5}$Si$_{3}$ would be a first essential step to elucidate the microscopic mechanisms responsible for the results reported so far and to investigate the response of the ordered and non-ordered Mn sites under this thermodynamic parameter.

\section{Conclusions}

In this study, we investigated the evolution of the spin-excitation spectrum of Mn$_{5}$Si$_{3}$ under external parameters.
The changes of the spin structure stimulated by temperature and magnetic field are also traced in the spin-excitation spectrum of Mn$_{5}$Si$_{3}$.
At low energies, all AFM phases host spin-wave excitations in agreement with the theoretical calculations. 
The temperature or field dependence of the spin gap allows a tracking of the changes in the spin-excitation spectrum between the AFM1 and AFM2 phases.
Apart from the spin gap (single vs double), the high-energy spin dynamics is also a fingerprint of each magnetic phase.
At high energies, whatever the combination of temperature and applied field, the spin dynamics of the AFM1 / AFM1' and AFM2 phases are very robust at consisting of either spin waves or broad fluctuations. 
The broad fluctuations pattern of the AFM2 phase at this stage cannot be addressed by our theoretical framework where on the magnetic sites (Mn1 and/or two-third of Mn2 sites) rigid moments are assumed.
Solving this puzzle likely requires a theory that extends the Heisenberg model to account for low-energy longitudinal spin fluctuations of the nominally non-magnetic Mn sites.
While the present overview gives a clear picture of the spin-excitation spectrum of Mn$_{5}$Si$_{3}$, the precise microscopic relationship between the peculiar spin dynamics and the evolution of the magnetically ordered and non-ordered Mn sites across the $H-T$ phase diagram is still a topic to be experimentally and theoretically addressed.

\section{Acknowledgements}

N.B.\ acknowledges the support of JCNS through the Tasso Springer fellowship and the Czech Science Foundation GA\v CR under the Junior Star Grant No. 21-24965M (MaMBA). 
F.J.d.S.\ acknowledges support of the European H2020 Intersect project (Grant No. 814487), and N.M.\ of the Swiss National Science Foundation (SNSF) through its National Centre of Competence in Research (NCCR) MARVEL.
The work of M.d.S.D. made use of computational support by CoSeC, the Computational Science Centre for Research Communities, through CCP9.
This work was also supported by the Brazilian funding agency CAPES under Project No.\ 13703/13-7 and the Deutsche Forschungsgemeinschaft (DFG) through SPP 2137 ``Skyrmionics'' (Project LO 1659/8-1).
We gratefully acknowledge the computing time granted through JARA on the supercomputer JURECA~\cite{jureca} at Forschungszentrum J\"ulich GmbH and by RWTH Aachen University.

\section{Authors declarations} 
%\subsection{Conflict of Interest}

The authors have no conflicts to disclose.

%\subsection{Author Contributions}

%\textcolor{red}{To be written...}

\section{Data availability} 

The neutron data collected at the ILL are available at Refs.~\onlinecite{data_Thales,data_IN12}.
The theoretical results  that support the findings of this study are available from the corresponding authors upon reasonable request.

\bibliography{paper_Mn5Si3_3}

\end{document}